\documentclass[fleqn,10pt]{PARCTwoColumn} 

\usepackage{lipsum} 
\usepackage{todonotes}
\usepackage[hyphenbreaks]{breakurl}
\usepackage[hyphens]{url}
\usepackage{gb4e}
\usepackage{framed}

\definecolor{color1}{rgb}{0,0,90} 
\definecolor{color2}{rgb}{0,20,20} 

\usepackage{algpseudocode}

\Masthead{networking and distributed systems\\computing science laboratory}

\PaperTitle{Process Migration over CCNx}

\Authors{Marc Mosko*\textsuperscript{1}} 
\affiliation{\textsuperscript{1}\textit{Computing Science Laboratory, PARC}} 
\affiliation{*\textbf{corresponding author}: marc.mosko@parc.com} 

\newcommand{\keywordname}{keywords} 

\Abstract{Process migration involves moving the running state of a process from one physical
system to another, as is commonly done for virtual machines.  In this paper, we describe how
Content Centric Networking (CCNx) facilitates process migration through an intuitive naming
ontology and version checkpointing.}

\begin{document}
\sloppy

\flushbottom 

\maketitle 

\tableofcontents


\thispagestyle{empty} 


\section{Introduction}
Process migration, or more generally virtual machine (VM) migration, is a method by which a running process (or VM) is moved
from one  system to another system.  It also applies to other processes, such as virtualized network functions [cite] or system slices [cite ITU Y-3000 series].
This paper describes how Content Centric Networking (CCNx) facilitates process migration while enabling many desirable
features such as strong checkpointing and data de-duplication.  Not all migration techniques require strong checkpointing, and
in those cases CCNx offers a faster and weaker naming technique that allows pages or blocks to go dirty in a checkpoint.

A standard method for VM migration is the \emph{pre-copy} method~\cite{clark2005live, Theimer:1985:PRE:323647.323629,travostino2006seamless}.
\emph{Pre-copy} divides migration in to three phases: push phase, stop-and-copy phase, and pull phase.  In the push phase, slowly changing state is moved over the network, possibly in several rounds.  In the stop-and-copy phase, the VM freezes and the hot state moves over the network.  In the pull phase, any remaining data not yet copied is page faulted over the network.  We assume there is a control channel between the VM source and the VM destination, as described in Section~\ref{sec:control}.  Travostino \emph{et al.}~\cite{travostino2006seamless} show the feasibility of using \emph{pre-copy} migration over high latency wide area networks.
For efficient \emph{pre-copy} migration, the migration agent needs to know the memory hot-spots, using techniques such as in Wood \emph{et al.}~\cite{wood2007black}.

The \emph{pre-copy} method is not the only approach, as the name would suggest.  Hines and Gopalan~\cite{hines2009post} show a method using \emph{post-copy} that is competitive with \emph{pre-copy}.  In a \emph{post-copy} approach, the CPU state is transferred first (the stop-and-copy phase above), and then the memory is
transferred, possibly page faulting some memory over the network after the VM is re-started.  While the present paper describes using CCNx in a \emph{pre-copy} mode, it should be clear how these same techniques can apply to \emph{post-copy} or other methods.

The Remus~\cite{cully2008remus} hot migration technique uses consistent and frequent checkpoints for hot-spare VM migration.  One or more second systems maintain near-real-time replicas of a primary and can take over computation if the primary fails.

There is a supervisory process, which is out of scope for the present work, that determines the need to move a process from one system to another system.  It instantiates a process duplicator agent on the source and destination systems, such as in the VM hypervisor.

In CCNx, a named address is the tuple \{CCNxName, KeyIdRest, HashRestr \}, where CCNxName is a hierarchical name (like a routable URI), KeyIdRestr is a restriction on the public key used to verify the cryptographic signature of the response, and HashRestr is a restriction on the SHA-256 hash of the response.  Inside the network, CCNx will ensure that the response carries the same CCNxName, the same KeyId as the KeyIdRestr, and the computed message hash equals the HashRestr.  The computed message hash is over the message body, which excludes some per-hop headers that could change without affecting the hash value.

The remainder of the paper is organized as follows.  Section~\ref{sec:transport} describes the transport features needed to operate over CCNx.
Section~\ref{sec:model} presents an example machine model and naming ontology, showing how using a CCNx approach naturally maps the machine model to network and storage resources.  Section~\ref{sec:migration} outlines how the migration process works and points out important events in each phase of a migration.
Section~\ref{sec:hashname} describes how using hash-based names and manifests allows for efficient checkpoint representation and transfer.
Section~\ref{sec:dedupe} describes how using CCNx with hash-based names and manifest leads to natural data de-duplication within a VM, between VMs, and even between VMs on different hosts.  Section~\ref{sec:routing} explains how routing interacts with the migration process.
Section~\ref{sec:control} describes the overall control channel used to orchestrate a migration and control the source and destination migration agents.
Finally, Section~\ref{sec:conclusion} concludes the paper.

\section{Data Transport in CCNx}\label{sec:transport}
We assume that the agents have a reliable transfer method between them.  We assume that the source and destination can agree on a window size or acknowledgement mechanism so the source and release resources correctly transferred to the destination.   We also assume there exists a close mechanism to release the final resources in the final window.  For example, the destination agent that is fetching data from the source using Interests has a method to re-transmit Interests.  It should also have a method to signal the source agent when it is done transferring a checkpoint so the source agent can release the memory and resources.  For example, when fetching a checkpoint prefixed by \url{/vm-name/checkpoint/ver=7}, it can fetch a virtual object named \url{vm-name-checkpoint/ver=7/close} indicating it wishes to terminate the transfer (i.e. it has received all the data its wants).  The source responds with an ACK Content Object.  The destination sends a final Interest for the virtual name \url{/vm-name/checkpoint/ver=7/close-ack}, to which the source sends another ACK Content Object.  This three-way (really four-way) handshake correctly terminates the session and the source knows it can now release its resources.

Having a reliable close mechanism is important in some cases because some actions might be dependent on both sides agreeing that a transfer is complete.  For example, if the source and destination will both use the same name prefix \url{/vm-name}, then they must agree when the change will take place, such as after finishing the stop-and-copy phase of data transfer.

\section{Machine Model}\label{sec:model}

A classic machine model is a central processing unit with a register file, random access memory, permanent storage (disk),
and accessories, such as a network interface or graphics system.  This leads to a fairly simple CCNx name mapping of a virtual
machine with a config file and then Content Objects that map to the specific system architecture.  In this example, the config file
would specify the hardware parameters, such as number of CPUs ($cpu_n$), the amount of RAM and page size (e.g. 1GB with 4KB pages), the number hard disks (e.g. hda and hdb), and network interfaces (e.g. en0).  Hard disks, for example, might be represented in the \emph{vhd}~\cite{vhd} format and have their own configuration file in addition to data blocks (e.g. 512 bytes or 4KB, as per the config file).

A standard disk model, such as \emph{vhd}~\cite{vhd} , uses three control structures (dynamic disk header, block allocation table (BAT), disk footer) in addition to the 512 byte data blocks.  

CCNx allows each of these machine elements to be represented in an intuitive way via hierarchical names.  In the example shown in Fig.~\ref{fig:naming}, we have used a fairly verbose hierarchy for sake of clarity.  In an actual implementation, the names could be shorter and some hierarchy levels compressed, if desired.

\begin{figure}[h]
\begin{verbatim}/vm-name/config
/vm-name/cpu/\{0 ... cpu_n\}/regfile
/vm-name/cpu/\{0 ... cpu_n\}/tlb}
/vm-name/ram/page/{0 ... ram_n}
/vm-name/disk/hda/config
/vm-name/disk/hda/vhd/header
/vm-name/disk/hda/vhd/bat
/vm-name/disk/hda/vhd/footer
/vm-name/disk/hda/block/{0 ... hda_n}
/vm-name/disk/hdb/config
/vm-name/disk/hdb/block/{0 ... hdb_n}
/vm-name/net/en0}
\end{verbatim}
\caption{Example virtual machine naming hierarchy}
\label{fig:naming}
\end{figure}

For example, let's model a virtual machine with 2 CPUs, one 2 GB hard disk that is 25\% full with 512 byte blocks, and 2GB of memory with 4KB pages.  The hard disk  uses 976,563 blocks plus 3 control structures plus config for a total of 976,567 Content Objects.
The memory uses 524,288 pages.  All told, this model uses 1,500,854 Content Objects.

\section{Migration}\label{sec:migration}
A standard method for VM migration is the \emph{pre-copy} method~\cite{clark2005live, Theimer:1985:PRE:323647.323629}.
\emph{Pre-copy} divides migration in to three phases: push phase, stop-and-copy phase, and pull phase.  In the push phase, slowly changing state is moved over the network, possibly in several rounds.  In the stop-and-copy phase, the VM freezes and the hot state moves over the network.  In the pull phase, any remaining data not yet copied is page faulted over the network.  We assume there is a control channel between the VM source and the VM destination, as described in Section~\ref{sec:control}.

In the push phase, the migration process decides which pieces of the system to checkpoint and creates uniquely named Content Objects for those.  A uniquely named Content Object has, for instance, a version number so it belongs to a consistent set of data.  In some cases, we might also use a hash-based name (see Section~\ref{sec:hashname}.  For example, the migration process might name these objects as \url{/vm-name/checkpoint/ver=j/chunk=k}, where $j$ is the version number of the checkpoint (e.g. $j = 0, 1, \dots$) and $k$ is a sequential number for each content object in that checkpoint.

The migration process needs to indicate the purpose of each chunk in a name like  \url{/vm-name/checkpoint/ver=j/chunk=k}.  This could be done by adding a metadata field that indicates its machine model placement (e.g. RAM page 3, HDA block 2), or it could be done via a CCNx Link.  Using a link, the chunk name would point to a spelled-out content name, such as \url{/vm-name/checkpoint/ver=j/ram/page/3}, so it is obvious where the data goes.  Using a link does not imply multiple round trips, as the link can be pre-pended to the fully named object.  These methods, however, are inefficient compared to the hash-based name method described in Section~\ref{sec:hashname}.

\begin{itemize}
\item The migration system iterates through push phases, creating sequential checkpoints and updating the destination system.
\item The stop-and-copy phase freezes the VM at the source, creates a next checkpoint of those critical resources, and transfers them to the destination.
\item In the pull phase, the source creates a final checkpoint of all remaining resources and the destination pulls them on-demand or at a leisurely pace.  Once the destination indicates to the source it has finished the final checkpoint transfer, the source and release all remaining resources.
\end{itemize}

\section{Using Nameless objects and Hash Names}\label{sec:hashname}
Nameless objects allow the system to:

\begin{itemize}
\item Store all data of the live system in Content Object memory -- disk blocks made of Content Objects and memory pages made of Content Objects.  This is because using Nameless Objects adds almost zero overhead, and the overhead it adds is usually constant.  For example, a 4KB memory region would have a 16-byte constant header (8 byte fixed header, 4 byte TLV open the T\_OBJECT and 4 byte TLV opening the T\_PAYLOAD).
\item The system can easily compute hash based names by hashing the fixed 8 bytes of TLV plus the data, e.g. 4KB page or 512 byte disk block.
\item The system does not need to generate a unique name for each item -- that is done by the hash based name.
\item The checkpoint now consists of a Manifest tree, where manifest entries use the hash based names of each item.
\item Once a resource has been hash and included in the manifest, the system uses something like a Copy-On-Write approach to maintain the checkpoint.
\item The root manifest contains metadata indicating its phase of transfer (push phase, stop-and-copy phase, pull phase)
\end{itemize}

A manifest set for a checkpoint would use a name prefix such as \url{/vm-name/checkpoint/ver=j/manifest/chunk=k}.  The checkpointed resources would use \emph{virtual} names such as \{\url{/vm-name/checkpoint/ver=j/ram}, hash=0x123\dots \}, where all the RAM pages included in the checkpoint have the same name prefix but only differ in hash value.  Inside the manifest, a page number would indicate where to put the page.  Similarly, for disk blocks the name prefixes are the same and only the hash values differ so duplicate blocks will not result in duplicate communications or storage.

\subsection{Weak Checkpoints}\label{sec:weak}
In some migration processes, it is not necessary to maintain a strong, consistent checkpoint.  For example, if the destination transfers a page of memory and
the source updates that page during or after the transfer, the source will mark the page to be sent in a later push phase.   The destination does not have
a consistent set, but that is acceptable because it is not using that set until after the stop-and-copy phase.

When using a weak checkpoint, those weak elements must not use a hash based name. Instead, those manifest sections would use a notation such as enumerating the RAM pages by page number.

\section{De-duplicating data}\label{sec:dedupe}
De-duplication is a technique where only one copy of data exists and it is shared between multiple instances. 
CCNx allows resources to be de-duplicated both \emph{within} and \emph{between} virtual machine instances.
For example, in the previous discussion about using hash names for resources, if two disk blocks, for example, have the same hash value they will refer to the same Content Object.  Only the block index in the manifest will be different.

A VM hypervisor may also share blocks between VMs.  When generating the names used to fetch a checkpoint, the source migration agent running in the source hypervisor could use a name like \{ \url{/nyc/host7}, hash = 0x63223\dots \} so any instance or any component can share the same data.  Assume that the memory page size and the disk block size are the same.  Then that name for hash 0x63223\dots could be both a disk block and a RAM page of the same data (e.g. a shared library code section).  Because the manifest can point to different name prefixes for each hash and can indicate the virtual resource of that hash, we can have the same physical bytes used for many purposes.

A migration agent could know, for example, that some disk blocks are common.  For example, \emph{hda} could mount a read-only root file system that only contains common OS and application binaries.  These could come from a name like \url{/nyc/objectstore} and be shared over many different physical hosts.  Using a manifest representation of a checkpoint allows those resources to come from that specific prefix while other resources come from host specific or vm specific locations.

The prior description illustrates how even a na\"{\i}ve  de-duplication approach is relatively simple and efficient in CCNx.  One can also use more
advanced de-duplication techniques that operate on smaller segments of data to achieve de-duplication even when, for example, the disk block size
does not match the memory page size.

\section{Routing}\label{sec:routing}
Routing may be managed several ways.  We assume that all systems have a unique name (e.g. \url{/nyc/host7/vm-name}, possibly in addition to a generic name (e.g. \url{/vm-name}).  We discuss how the migration process works in each of these models:
\begin{itemize}
\item External: An external agency or agencies manage the routing namespace, for example today's Internet.
\item Software Defined: A central, but programmable, agency manages the routing namespace, for example an SDN environment.
\item Distributed: The endpoints manage the routing namespace, such as by running a secure routing process.
\end{itemize}

In the External model, the source and destination migration agents will have different names.  The source might have the name prefix \url{/nyc/host7/vm-name}
and the destination \url{/sfo/host2/vm-name}.  The migration orchestrator would understand these names and appropriately instruct the migration agents at each site of the correct names.

In a software-defined model, it is possible to use generic names such as \url{/vm-name}.  Prior to and during the stop-and-copy phase, the name points to the source agent.  After the stop-and-copy phase, when the destination is ready to start the VM, it notifies the network controller to point \url{vm-name} to the destination host.  The source host now only has its location-dependent name, such as \url{/nyc/host7/vm-name}, which is used in the pull phase to transfer any remaining data.

In the distributed model, the source agent advertises \url{/vm-name} until the completion of the stop-and-copy phase.  After this point, it stops advertising the name and the destination agent begins advertising it.  The destination agent may now finish transferring data in the pull phase using the location-dependent name of the source agent.

There are other possible solutions, which could result in equally correct behavior.

\section{Control Channel}\label{sec:control}

We assume that CCNx routing is setup such that \url{/vm-name} points to the correct location of the source system.  The destination agent will poll the source agent until the source agent is running, and will then request the first checkpoint, such as by issuing an Interest for \url{/vm-name/checkpoint/ver=0/manifest}.  After a specific checkpoint version is transferred, the destination agent can pull the next checkpoint version.

Each checkpoint version manifest, for example, describes the purpose of that checkpoint.  While receiving push phase manifests, the destination agent knows to keep the VM frozen.  After it receives the stop-and-copy phase manifest, it can start the VM, at which point it would also want to retrieve the final manifest for the next checkpoint version which is the remaining un-copied data that the destination agent can pull at its leisure.

\section{Example Migration Process}\label{sec:example}
This section walks through a complete example of transferring a machine.  These examples do not cover security aspects of the migration, though we note that the supervisory process can mediate tokens or credentials between the parties and existing key exchange and security protocols can provide high-speed, authenticated, and private communications.

\subsection{Manifests with nameless objects}
\begin{enumerate}
\item A supervisory process identifies a source \url{/parc/vm3} to be transferred to a destination \url{/parc/vm7}. 
\item The destination polls the source with an interest for \url{/parc/vm3/checkpoint/ver=0/manifest}, where it knows to always start at version 0~\footnote{The supervisory process could provide a base name and version number to avoid always starting with the same name.}.
\item The source builds the version 0 checkpoint, for example using techniques form Clark~\cite{clark2005live} to prioritize low-turnover memory pages and disk blocks.  For example, \emph{VM direct memory} (memory paged in to the kernel) is considered hot and paged out memory is used in the push phase.
	\begin{enumerate}
   	\item One section of the manifest identifies this as a \emph{pre-copy} phase transfer.
	\item One section points to Vm configuration files, such as the overall config and disk descriptors.
	\item One section identifies the following links as part of RAM pages and identifies the page in the manifest and points to a hash-based name.
	\item One section identifies the following links as part of disk blocks and identifies the block number in the manifest and points to a hash-based name.
	\end{enumerate}
\item The source publishes the root manifest of the checkpoint, answering the Interest from the destination, which begins transferring that checkpoint.
\item The source repeats this processes until the marginal gain of doing \emph{pre-copy} phases becomes small.
\item The source freezes the VM and creates a checkpoint that covers all important resources for operation of the VM, such as CPU state and high-turnover pages.
	\begin{enumerate}
   	\item One section of the manifest identifies this as a \emph{stop-and-copy} phase transfer.
	\end{enumerate}
\item The destination copies receives the \emph{stop-and-copy} phase manifest and data, then starts the VM.
\item After the destination indicates that it is done, the source may release all copied resources.
\item The source creates a final checkpoint of any remaining uncopied resources that the destination can lazily pull as needed.
\item After the destination finishes copying the last checkpoint, the source can free remaining resources.
\end{enumerate}

\section{Conclusion}\label{sec:conclusion}
We have described how CCNx elegantly and efficiently solves the process migration problem using
intuitive naming and strong checkpointing for correct and safe data transfer.  We modeled a typical
machine architecture and used a common \emph{pre-copy} transfer scheme, though other models
and schemes could be employed.  Using hash based names and manifests, we showed how CCNx
enables several desirable features, such as data de-duplication \emph{within a VM}, \emph{between VMs},
and \emph{between physical systems}.  In fact, by hash-naming the network resources and
indicating their use in a manifest, we can even share de-duplicated blocks between disk images and
RAM pages.


\bibliographystyle{unsrt}
\bibliography{ref}


\end{document}